\documentclass[aps,preprint,showpacs,groupedaddress,floatfix]{revtex4}
\usepackage{graphicx}
\usepackage{dcolumn}
\usepackage{bm}

\newcommand{\lp}{\left}
\newcommand{\rp}{\right}

\begin{document}

\title{Inclusive charged-current neutrino-nucleus reactions 
calculated with the relativistic quasiparticle random phase approximation}
\author{N. Paar}
\author{D. Vretenar}
\author{T. Marketin}
\affiliation{Physics Department, Faculty of Science, University of Zagreb, 
Croatia}
\author{P. Ring}
\affiliation{Physik-Department der Technischen Universit\"at M\"unchen, 
D-85748 Garching,
Germany}
\date{\today}

\begin{abstract}
Inclusive neutrino-nucleus cross sections are calculated using a 
consistent relativistic mean-field theoretical framework. The weak 
lepton-hadron interaction is expressed in the standard 
current-current form, the nuclear ground state is described with the 
relativistic Hartree-Bogoliubov model, and the relevant transitions to 
excited nuclear states are calculated in the relativistic quasiparticle random 
phase approximation. Illustrative test calculations are performed for
charged-current neutrino reactions  on $^{12}$C, $^{16}$O, 
$^{56}$Fe, and $^{208}$Pb, and 
results compared with previous studies and available data.
Using the experimental neutrino fluxes, the averaged 
cross sections are evaluated for nuclei of interest for neutrino
detectors. We analyze the total neutrino-nucleus cross sections, 
and the evolution of the contribution of the different  multipole
excitations as a function of neutrino energy.  
The cross sections for reactions of supernova neutrinos
on $^{16}$O and $^{208}$Pb target nuclei are analyzed 
as functions of the temperature and chemical potential.
 
\end{abstract}

\maketitle

\bigskip \bigskip
%%%%%%%%%%%%%%%%%%%%%%%%%%%%%%%%%%%%%%%%%%%%%%%%%%%%%%%%%%%%%%%%%%%%%%%%%%
%  Section 1
\section{\label{introduction}Introduction}
%%%%%%%%%%%%%%%%%%%%%%%%%%%%%%%%%%%%%%%%%%%%%%%%%%%%%%%%%%%%%%%%%%%%%%%%%%

Neutrino-nucleus reactions at low energies play an important role in many 
phenomena in nuclear and particle physics, as well as astrophysics. These reactions 
present extremely subtle physical processes, not only 
because they involve the weak interaction, but also because they are very 
sensitive to the structure of nuclear ground states and excitations, 
i.e. to the solution of the nuclear many-body problem that includes the 
strong and electromagnetic interactions. The use of microscopic nuclear 
structure models in a consistent theoretical framework is therefore essential 
for a quantitative description of neutrino-nucleus reactions~\cite{Hay.99}. 
Detailed predictions of neutrino-nucleus cross sections are crucial for the
interpretation of neutrino experiments, detection of neutrinos produced in
supernova explosions and understanding the underlying nature of
these explosions~\cite{sns1}.
Neutrino-nucleus reactions which occur in 
a type II supernova could also contribute to the nucleosynthesis~\cite{Heg.05,Lan.06},
but more data on cross sections are necessary for a more complete
understanding of this process, as well as the supernova dynamics.    

Data on neutrino-nucleus cross sections have been obtained by the
LSND~\cite{Alb.95,Ath.97} and KARMEN~\cite{Bod.92,Bod.94,Mas.98} 
collaborations, and at LAMPF~\cite{Kra.92,Koe.92}, but only for 
$^{12}$C and $^{56}$Fe target nuclei. New experimental programs are being 
planned which will provide essential data on the neutrino-nucleus reactions, 
and also help to improve the reliability of present cross-section calculations.
These include the spallation neutron source (SNS) at ORNL, where neutrinos 
produced by pion decay at rest will enable measurements of cross
sections for a wide range of target nuclei~\cite{Avi.03,Efr.05}, and the
promising ``beta-beams'' method for the production of pure electron 
neutrino-beams by using the $\beta$-decay of boosted 
radioactive ions~\cite{Zuc.02,Vol.04}. Cross sections for neutrino energies
in the range of tens of MeV could be measured, and these reactions  
are particularly interesting for supernova studies~\cite{McL.04}.

Weak interaction rates at low energies have been analyzed employing a variety of 
microscopic approaches, principally in the frameworks of the shell
model~\cite{Hax.87,Eng.96,Hay.00}, the random phase 
approximation (RPA)~\cite{Aue.97,Sin.98,Vol.00,Vol.02}, 
continuum RPA (CRPA)~\cite{Kol.92,Kol.95,Jac.99,Jac.02,Bot.05},
hybrid models of CRPA and shell model~\cite{Kol.99,Kol.03},
and the Fermi gas model~\cite{Wal.75,Gai.86,Kur.90}.  
The shell model provides a very accurate description of ground state 
wave functions. The description of high-lying excitations, however, 
necessitates the use of large model spaces and this often 
leads to computational difficulties, making the approach applicable 
essentially only to light and medium-mass nuclei. For systematic 
studies of weak interaction rates throughout the nuclide chart, microscopic 
calculations must therefore be performed using models based on the RPA.

 Hybrid models combine the shell-model and CRPA in such a way that occupation
probabilities of single-particle states in a specific nucleus are determined
by shell model calculations, and then inserted into the CRPA~\cite{Kol.99}. 
In general the CRPA employs different interactions for the calculation of the 
nuclear ground state (for instance, the Woods-Saxon potential),
and in the residual CRPA interaction (G matrix from the Bonn potential, or the
Landau-Migdal interaction), and thus additional parameters are required in order 
to adjust the calculated rates to data.
A more consistent approach, based on the quasiparticle RPA with Skyrme effective
interactions, has been employed in calculations of weak interaction 
rates~\cite{Vol.00}. Although in this framework the residual
RPA interaction is derived from the same energy functional which determines
the nuclear ground state, some terms are usually omitted, and 
pairing correlations require the adjustment of additional factors. A fully consistent theoretical framework for the description of charge-exchange excitations in open 
shell nuclei, based on Skyrme effective interactions, has only recently been 
developed  \cite{Fra.05}, but not yet employed in the analysis of neutrino-nucleus 
reactions. Neutrino-nucleus cross sections have also 
been calculated using the ab-initio no-core shell model based on realistic NN
and three-body interactions~\cite{Hay.03}, and the shell model with an improved
Hamiltonian which properly takes into account the spin-isospin
interactions~\cite{Suz.06}. The importance of improved calculations of 
neutrino-nucleus cross sections for neutrino-oscillation studies 
has been demonstrated in the recent reanalysis of the LSND 
experiment~\cite{Sam.06}, using the particle-number 
projected quasiparticle RPA~\cite{Krm.05}, which has shown an 
enhancement of the neutrino-oscillation probability when compared to 
previous studies.

Although the general expressions for the transition matrix elements 
relevant for the calculation of neutrino-nucleus cross sections have been 
known since many years~\cite{Con.72,Don.80}, it is only more recently that
systematic calculations have been performed in open-shell nuclei by 
making use of modern effective interactions in the description of both nuclear
ground states and excitations~\cite{Vol.00}. Reliable prediction of weak
interaction rates in nuclei necessitates a fully consistent description 
of the structure of ground states and multipole excitations.
Among the relevant charge-exchange excitations, 
the isobaric analog state (IAS) and Gamow-Teller resonance 
(GTR) have been the subject of extensive experimental and 
theoretical studies. Much more limited are the data and 
theoretical predictions for properties of excitations of 
higher multipolarities at finite momentum transfer.

In this work we analyze charged-current neutrino-nucleus reactions by 
employing a fully consistent microscopic approach based on relativistic 
energy density functionals, and also including pairing correlations in 
the description of open-shell target nuclei. An essential advantage
over most current approaches is the use of a single universal effective
interaction in calculations of both ground-state properties and 
multipole excitations of nuclei in various mass regions of the chart 
of nuclides.

Of particular interest for the present study are rates for neutrino-nucleus
reactions in the low-energy range below 100 MeV, 
which play an important role in many astrophysical processes, including
stellar nucleosynthesis. A quantitative description of nucleosynthesis
of heavy elements during the r-process necessitates accurate predictions
of neutrino-nucleus cross sections not only in stable nuclei,
but also in nuclei away from the valley of $\beta$-stability. Since nuclei
are used as detectors for solar and supernovae neutrinos, as well as in neutrino
oscillation experiments, it is important to describe
the neutrino detector response in a consistent and fully 
microscopic theory. Finally, a quantitative estimate of neutrino-nucleus
reaction rates will provide information relevant for feasibility studies and
simulations of a low-energy beta beam facility, which could be used to
produce neutrino beams of interest for particle physics, nuclear physics
and astrophysics~\cite{Vol.04}.

In Sec.~\ref{sectionII} we outline the basic formalism
used in the evaluation of neutrino-nucleus cross sections. In 
Sec.~\ref{sectionIII} the relativistic Hartree-Bogoliubov 
model (RHB), and the proton-neutron relativistic quasiparticle random phase 
approximation (PN-RQRPA) are briefly reviewed. 
Section~\ref{sectionIV} includes several test cases, 
and the calculated cross sections are compared with results of 
previous theoretical studies and the available data. Cross sections 
for supernova neutrinos are analyzed in Sec.~\ref{sectionV}, and finally
Sec.~\ref{sectionVI} summarizes the results of the present investigation 
and ends with an outlook for future studies. 

%----------------------------------------------------------------------------
\section{\label{sectionII}Neutrino-nucleus cross sections}
%----------------------------------------------------------------------------

In the present study we consider the charged-current neutrino-nucleus reactions:
\begin{equation}
\nu_l + _{Z} X _N  \rightarrow _{Z+1}X^*_{N-1} + l^- \;,
\end{equation}
where $l$ denotes the charged lepton (electron,  muon).
Detailed expressions for the reaction rates and the transition
matrix elements can be found in Refs.~\cite{Con.72,Wal.75}.
The charged-current neutrino-nucleus cross section reads
\begin{equation}
\label{CS}
\left (  \frac{d \sigma_{ \nu  }   }{ d \Omega }   \right ) =
 \frac{1 }{ (2 \pi )^2}V^2 p_l E_l
 \sum_{\begin{minipage}{0.7cm} \scriptsize lepton \\  spins \end{minipage} } \frac{ 1}{ 2J_i + 1}   \sum_{M_i M_f} | \langle
f |  \hat{H}_{W} |i  \rangle | ^2 \;,
\end{equation}
where $p_l$ and $E_l$ are the momentum and energy of the outgoing lepton,
respectively. The Hamiltonian $ \hat{H}_{W} $ of the weak interaction 
is expressed in the standard current-current form, i.e. in terms of the 
nucleon $\mathcal{J}_{ \lambda }(\bm{x})$ and lepton $j_{ \lambda }(\bm{x})$ 
currents
\begin{equation}
\hat{H}_{W}=- \frac{G }{ \sqrt{2} }  \int d\bm{x}
\mathcal{J}_{ \lambda }(\bm{x})j^{ \lambda }(\bm{x}) \;,
\end{equation}
and the transition matrix elements read
\begin{equation}
\langle f | \hat{H}_{W}| i \rangle = -  \frac{ G}{\sqrt{2}}l_{ \lambda }
 \int d \bm{x} e^{- i \bm{q} \bm{x}}
\langle f | \mathcal{J}^{ \lambda }(\bm{x}) | i \rangle \; .
\end{equation}
The multipole expansion of the leptonic matrix element 
$l_{ \lambda }  e^{- i \bm{q} \bm{x}}$ determines 
the operator structure for the nuclear transition matrix elements~\cite{Wal.75}, 
and the expression for the neutrino-nucleus cross section. In the 
{\em extreme relativistic limit|} (ERL), in which the energy of
the outgoing lepton is considered much larger than its rest mass, 
the differential neutrino-nucleus cross section takes the form
\begin{eqnarray}
& &\left (  \frac{d \sigma_{ \nu  }   }{ d \Omega }   \right )_{ERL}
  =  \frac{ 2G_F^2cos^2 \theta_c }{\pi}  
\frac{E_l^2}{ 2J_i+1} \nonumber \\
& &\times  \Bigg\{ {\left( \frac{ q^2}{ 2\bm{q}^2}cos^2  \frac{  \theta }{2} +sin^2 \frac{ \theta  }{2 }   \right )}
\sum_{ J \geq 1} \left[ \vert \langle J_f || \hat{\mathcal{T}}_J^{MAG} || J_i \rangle \vert^2 \ 
+ \vert \langle J_f || \hat{\mathcal{T}}_J^{EL} || J_i \rangle \vert^2 \ \right ] \nonumber \\
& & { -sin \frac{ \theta  }{2 } \sqrt{  \frac{q^2}{\bm{q}^2} 
cos^2 \frac{ \theta  }{ 2}+ sin^2 \frac{ \theta  }{2}}}  
\sum_{J \geq 0 } 2Re\langle J_f || \hat{\mathcal{T}}_J^{MAG} || J_i \rangle 
\langle J_f || \hat{\mathcal{T}}_J^{EL}|| J_i \rangle^{*} \nonumber \\
& & {+cos^2 \frac{  \theta }{2 }}  \sum_{J  \geq 0 }
|  \langle J_f || \hat{\mathcal{M}}_J 
- \frac{q_0}{|\bm{q}| } \hat{\mathcal{L}}_J || J_i \rangle | ^2 \Bigg\}.
\label{ncs_eq}
\end{eqnarray}
where $G_F$ is the Fermi constant for the weak interaction,  $\theta_c$ is the
Cabbibo's angle, $\theta$ denotes the angle between the incoming and outgoing leptons,
the energy of the lepton in the final state is $E_l$, 
and the 4-momentum transfer is $q \equiv (q_0, \bm{q})$. The nuclear transition 
matrix elements between the initial state 
$|J_i \rangle$ and final state $|J_f \rangle$, correspond to the 
charge $\hat{\mathcal{M}}_J$, longitudinal $\hat{\mathcal{L}}_J$, 
transverse electric $ \hat{\mathcal{T}}_J^{EL}$, and 
transverse magnetic $\hat{\mathcal{T}}_J^{MAG}$ multipole operators. 
These are expressed in terms of spherical Bessel functions, spherical harmonics, 
and vector spherical harmonics:
\begin{equation}
M_J^M \equiv j_J (\kappa x) Y_J^M ( \Omega_x)
\end{equation}
\begin{equation}
\bm{M}_{JL}^M \equiv
j_L (\kappa x) \bm{Y}_{JL1}^M ( \Omega_x) \; ,
\end{equation}
where we use the standard notation $\kappa = |\bm{q}|$.
There are four transition operators in Eq.~(\ref{ncs_eq}):
\begin{itemize}
\item the Coulomb multipole operator
\begin{equation}
\hat{\mathcal{M}}_{JM}(\bm{x})= F_1^V M_J^M(\bm{x})
- i \frac{ \kappa  }{m_N} 
\left[  F_A  \Omega _J^M(\bm{x})
+ \frac{ 1}{ 2}  (F_A-q_0 F_P)  \Sigma''^M_J(\bm{x})   \right ]  \;,
\label{coulombop}
\end{equation}
\item the longitudinal operator
\begin{equation}
\hat{\mathcal{L}}_{JM}(\bm{x})=  -\frac{q_0}{ \kappa }F_1^V M_J^M(\bm{x})
+i  \left[ F_A - \frac{ 1}{ 2}  \frac{  \kappa ^2 }{ m_N} F_P  \right ]  
\Sigma ''^M_J(\bm{x}) \;,
\label{longitudinalop}
\end{equation}
\item the transverse electric operator
\begin{equation}
\hat{\mathcal{T}}_{JM}^{el}(\bm{x})=  \frac{\kappa }{ m_N}
 \left[ F^V_1  {\Delta'}_{J}^{M}(\bm{x})+ \frac{1 }{2 } \mu^V
 \Sigma_J^M(\bm{x})     \right ]
+i F_A  {\Sigma'}_J^M (\bm{x}) \;,
\label{transverseelop}
\end{equation}
\item and the transverse magnetic operator
\begin{equation}
\hat{\mathcal{T}}_{JM}^{mag}(\bm{x}) = -i  \frac{ \kappa  }{m_N }
 \left[ F_1^V  \Delta_J^M(\bm{x})- \frac{1 }{2 }  \mu^V  {\Sigma'}_J^M(\bm{x})     \right ]
+F_A  \Sigma_J^M(\bm{x}) \; ,
\label{transversemagop}
\end{equation}
\end{itemize}
where all the form factors are functions of $q^2$.
These multipole operators contain seven fundamental
operators expressed in terms of  spherical Bessel functions,
spherical harmonics, and vector spherical harmonics~\cite{Con.72}:
\begin{equation}
M_J^M(\bm{x})
\end{equation}
\begin{equation}
 \Delta_J^M(\bm{x}) =  \bm{M}^M_{JJ} (\bm{x})
\frac{1 }{ \kappa  } \bm{\nabla}
\end{equation}
\begin{equation}
 {\Delta'}_J^M(\bm{x}) = \left[ -\sqrt{\frac{J}{2J+1}} \bm{M}^M_{J J+1}(\bm{x})
+\sqrt{\frac{J+1}{2J+1}} \bm{M}^M_{J J-1}(\bm{x}) \right]
 \frac{ 1}{  \kappa } \bm{ \nabla }
\end{equation}
\begin{equation}
{\Sigma}_J^M(\bm{x}) =\bm{M}^M_{J J}(\bm{x})\bm{\sigma}
\end{equation}
\begin{equation}
{\Sigma'}_J^M(\bm{x}) =
\left[ -\sqrt{\frac{J}{2J+1}} \bm{M}^M_{J J+1}(\bm{x})
+\sqrt{\frac{J+1}{2J+1}} \bm{M}^M_{J J-1}(\bm{x}) \right]
\bm{ \sigma  }
\end{equation}
\begin{equation}
{\Sigma''}_J^M(\bm{x}) =
\left[ \sqrt{\frac{J+1}{2J+1}} \bm{M}^M_{J J+1}(\bm{x})
+\sqrt{\frac{J}{2J+1}} \bm{M}^M_{J J-1}(\bm{x}) \right]
\bm{ \sigma  }
\end{equation}
\begin{equation}
 \Omega _J^M(\bm{x})=M_J^M(\bm{x}) \bm{\sigma} \frac{1 }{ \kappa  } \bm{ \nabla }.
\end{equation}
By assuming conserved vector current (CVC), the standard 
set of form factors reads \cite{Con.72,Kur.90}: 
\begin{equation}
F_1^V(q^2) =    \left ( 1+ \frac{q^2 }{(855 MeV)^2 }   \right )^{-2}
\end{equation}
\begin{equation}
\mu^V(q^2) =    4.706 \left ( 1+ \frac{q^2 }{(855 MeV)^2 }   \right )^{-2}
\end{equation}
\begin{equation}
F_A(q^2) =    -1.23 \left ( 1+ \frac{q^2 }{(855 MeV)^2 }   \right )^{-2}
\end{equation}
\begin{equation}
F_P(q^2) =  \frac{2 m_N F_A(q^2)}{q^2 +m_{ \pi }^2} \; .
\end{equation}
The neutrino-nucleus cross section is evaluated using Eq.~(\ref{ncs_eq}), with the transition matrix elements between the initial and final states determined in a
fully microscopic theoretical framework based on the Relativistic Hartree-Bogoliubov (RHB) model for the nuclear ground state, and excited states are calculated using the relativistic quasiparticle random phase approximation (RQRPA). 
More details about the RHB+RQRPA framework are presented in the next section.
%=========================================================================
\section{\label{sectionIII} The proton-neutron relativistic quasiparticle random phase approximation}
%-------------------------------------------------------------------------

Here we outline the essential features of the theoretical framework
that will be used to evaluate inclusive neutrino-nucleus cross sections: 
the proton-neutron (PN)-RQRPA based on relativistic Hartree-Bogoliubov model. 
The detailed formalism, as well as recent applications of these models 
in the description of exotic structure phenomena in nuclei far from 
stability, have been reviewed in Refs.~\cite{Vre.05,PVKC.07}.

The RQRPA has been formulated in the canonical single-nucleon basis of the 
relativistic Hartree-Bogoliubov (RHB) model in Ref.~\cite{Paar.03}, and extended
to the description of charge-exchange excitations (proton-neutron RQRPA) 
in Ref.~\cite{PNVR.04}. In the RHB framework the ground state of a nucleus 
can be written either in the quasiparticle basis as a product of independent 
quasi-particle states, or in the canonical basis as a highly correlated BCS-state. 
By definition, the canonical basis diagonalizes the density matrix and is always
localized. It describes both the bound states and the positive-energy single-particle
continuum. The QRPA equations in the canonical basis include the 
matrix elements $V_{\kappa ^{{}}\lambda ^{\prime }\kappa ^{\prime }\lambda
^{{}}}^{ph}$ of the residual particle-hole
($ph$) interaction, and the matrix elements $V_{\kappa
^{{}}\kappa ^{\prime }\lambda ^{{}}\lambda ^{\prime }}^{pp}$ 
of the particle-particle ($pp$) (pairing) interaction, as well as certain combinations 
of occupation factors $u_{\kappa }$, $v_{\kappa }$ of the canonical 
single-nucleon states. In addition to the configurations built from two-quasiparticle
states of positive energy, the relativistic QRPA configuration space must also include
pair-configurations formed from the fully or partially occupied states of
positive energy and the empty negative-energy states from the Dirac sea. The
inclusion of configurations built from occupied positive-energy states and
empty negative-energy states is essential for current conservation, 
decoupling of spurious states, and a quantitative description of excitation
energies of giant resonances~\cite{Paar.03}.  

The RHB+RQRPA model is fully consistent: in the particle-hole channel effective
Lagrangians with density-dependent meson-nucleon couplings are employed, 
and pairing correlations are described by the pairing part of the finite range Gogny
interaction~\cite{BGG.91}. Both in the $ph$ and $pp$ channels, the same interactions
are used in the RHB equations that determine the canonical quasiparticle basis, and 
in the matrix equations of the RQRPA. This is very important because the 
energy weighted sum rules are fulfilled only if the pairing interaction is
consistently included both in the static RHB and in the dynamical RQRPA
calculation. In the present work all RHB plus QRPA calculations have been 
performed using one of the modern and most accurate meson-exchange 
density-dependent relativistic mean-field effective interactions -- 
DD-ME2~\cite{LNVR.05} in the $ph$ channel, and the 
finite range Gogny interaction D1S~\cite{BGG.91} in the $pp$ channel.

Charge-exchange excitations have been analyzed with the
proton-neutron RQRPA using effective Lagrangians characterized by 
density-dependent meson-nucleon couplings~\cite{PNVR.04}.
The RHB plus PN-RQRPA framework has been successfully employed
in studies of isobaric analog resonances, Gamow-Teller resonances \cite{PNVR.04}, 
the relation between the excitation energies of IAS and GTR and the 
evolution of neutron skin in neutron-rich nuclei \cite{Vre.03}, and $\beta$-decay
rates of r-process nuclei~\cite{Nik.05,Mar.07}. The model includes both
the $T=1$ and $T=0$ pairing channels, and represents a relativistic extension
of the fully-consistent proton-neutron QRPA introduced in Ref.~\cite{Eng.99}.

The matrix equations of the PN-RQRPA read
\begin{equation} \left( \begin{array} [c]{cc}
A & B\\
B^{^{\ast}} & A^{^{\ast}}
\end{array}
\right)  \left( \begin{array} [c]{c}
X^{\lambda }\\
Y^{\lambda }
\end{array}
\right)  =E_{\lambda}\left( \begin{array} [c]{cc}
1 & 0\\
0 & -1
\end{array}
\right)  \left( \begin{array} [c]{c}
X^{\lambda }\\
Y^{\lambda }
\end{array}\right) \; , 
\label{pnrqrpaeq}
\end{equation}
with the matrices $A$ and $B$ defined in the canonical basis
\begin{eqnarray}
A_{pn,p^\prime n^\prime} &=& H^{11}_{pp^\prime}\delta_{nn^\prime} +
  H^{11}_{nn^\prime}\delta_{pp^\prime}  \nonumber \\ & & +
\lp( u_p v_n u_{p^\prime} v_{n^\prime} + v_p u_n v_{p^\prime} u_{n^\prime}\rp)
 V_{pn^\prime n p^\prime}^{ph } + 
\lp( u_p u_n u_{p^\prime} u_{n^\prime} + v_p v_n v_{p^\prime} v_{n^\prime}\rp) 
 V_{pn p^\prime n^\prime}^{pp } \nonumber \\
B_{pn,p^\prime n^\prime} &=& 
\lp( u_p v_n v_{p^\prime} u_{n^\prime} + v_p u_n u_{p^\prime} v_{n^\prime}\rp)
 V_{pp^\prime n n^\prime}^{ph } \nonumber \\ & &- 
\lp( u_p u_n v_{p^\prime} v_{n^\prime} + v_p v_n u_{p^\prime} u_{n^\prime}\rp) 
 V_{pn p^\prime n^\prime}^{pp } \; .
\label{abmat}
\end{eqnarray}
The proton and neutron quasiparticle canonical states are denoted 
by $p$, $p^\prime$, and $n$, $n^\prime$, respectively. 
$V^{ph}$ is the proton-neutron particle-hole residual 
interaction, and $V^{pp}$ denotes the particle-particle interaction.
The canonical basis diagonalizes the density matrix, and 
the corresponding eigenvalues are the occupation probabilities $v^2_{p(n)}$. 
Since on the other hand the canonical basis does not diagonalize 
the Dirac single-nucleon mean-field Hamiltonian $\hat{h}_{D}$, 
nor the pairing field $\hat{\Delta}$, the off-diagonal matrix elements 
$H^{11}_{nn^\prime}$ and $H^{11}_{pp^\prime}$ appear in Eq. (\ref{abmat}):
\begin{equation}
H_{\kappa \kappa^\prime}^{11}=(u_{\kappa }u_{\kappa^\prime }
-v_{\kappa }v_{\kappa^\prime
})h_{\kappa \kappa^\prime }-(u_{\kappa }v_{\kappa^\prime }+
v_{\kappa }u_{\kappa^\prime
})\Delta _{\kappa \kappa^\prime } \;.
\label{H11}
\end{equation}
The solution of the RQRPA matrix equation (\ref{pnrqrpaeq}) determines the 
excitation energies $E_{\lambda}$ and the corresponding forward- and 
backward-going amplitudes $X^{\lambda}$ and $Y^{\lambda}$, respectively. 
In the evaluation of the neutrino-nucleus cross section (Eq.~(\ref{ncs_eq})), 
for each transition operator $\hat{O}^J$ the matrix elements  
between the ground state of the even-even (N,Z) target nucleus 
and the final state are expressed in terms of single-particle matrix elements 
between quasiparticle canonical states, occupation probabilities, and
the corresponding RQRPA amplitudes:
\begin{equation}
 \langle J_f || \hat{O}^J || J_i \rangle =
 \sum_{pn} 
 \langle p || \hat{O}^J || n \rangle
\lp( X_{pn}^{J} u_p v_n - Y_{pn}^{J}v_p u_n \rp).
\label{redtrans}
\end{equation}
In the calculations performed in this work we have considered
transitions between the $|0^+\rangle$ ground state of a spherical even-even
target nucleus and excited states in the odd-odd nucleus with multipolarities 
$J_f^{\pi}=0^{\pm}$ -- $7^{\pm}$.

The spin-isospin-dependent interaction terms are generated by the 
exchange of $\pi$ and $\rho$ mesons. 
Although the direct one-pion contribution to the nuclear ground state 
vanishes at the mean-field level because of parity conservation, the 
pion nevertheless must be included in the calculation of spin-isospin 
excitations which contribute to the neutrino-nucleus cross section. 
The corresponding particle-hole residual interaction of the PN-RQRPA 
is derived from the Lagrangian density 
\begin{equation}
\mathcal{L}_{\pi + \rho}^{int} = 
      - g_\rho \bar{\psi}\gamma^{\mu}\vec{\rho}_\mu \vec{\tau} \psi 
      - \frac{f_\pi}{m_\pi}\bar{\psi}\gamma_5\gamma^{\mu}\partial_{\mu}
        \vec{\pi}\vec{\tau} \psi \; , 
\label{lagrres}	
\end{equation}
where arrows denote vectors in isospin space. The residual two-body interaction reads
\begin{eqnarray}
V(\bm{r}_1,\bm{r}_2) &=& \vec{\tau}_1\vec{\tau}_2 (\beta \gamma^\mu)_1
      (\beta \gamma_\mu)_2 g_\rho[\rho_v(\bm{r}_1)] g_\rho[\rho_v(\bm{r}_2)]
      D_\rho (\bm{r}_1,\bm{r}_2) \nonumber \\
      &-&\left(\frac{f_\pi}{m_\pi}\right)^2\vec{\tau}_1\vec{\tau}_2
      (\bm{\Sigma}_1\bm{\nabla}_1)(\bm{\Sigma}_2\bm{\nabla}_2)
      D_\pi (\bm{r}_1,\bm{r}_2)\;,
\end{eqnarray}
where $D_{\rho (\pi)}$ is the meson propagator
\begin{equation}
 D_{\rho (\pi)} = \frac{1}{4\pi}
        \frac{e^{-m_{\rho (\pi)}|\bm{r}_1-\bm{r}_2|}}{|\bm{r}_1-\bm{r}_2|}\;, 
\end{equation}	   
and
\begin{equation}
\bm{\Sigma} = \left(
\begin{array}
[c]{cc}%
\bm{\sigma} & 0\\
0 & \bm{\sigma}%
\end{array}
\right)  \;. \label{bigsigma}%
\end{equation}
%
%%%%%%%%%%%%%%%%%%%%%%%%%%%%%%%%%%%%%%%%%%%%%%%%%%%%%%%%%%%%%%%%%%%%%%%%%%%
For the $\rho$-meson density-dependent coupling strength we choose the functional
form used in the DD-ME2 effective interaction~\cite{LNVR.05}:
\begin{equation}
g_\rho (\rho_v) = g_\rho(\rho_{sat}){\rm e}^{-a_\rho (x-1)} \;,
\end{equation}
with $x=\rho_v / \rho_{sat}$, and $\rho_{sat}$ is the
nucleon density at saturation in symmetric nuclear matter.

For the pseudovector pion-nucleon coupling we use the standard values: 
$f_{\pi}^{2}/{4\pi}=0.08$, and $m_{\pi}=138$ MeV.
The derivative type of the pion-nucleon coupling necessitates the  
inclusion of the zero-range
Landau-Migdal term, which accounts for the contact part of the 
nucleon-nucleon interaction
\begin{equation}
V_{\delta\pi} = g^\prime \lp( \frac{f_\pi}{m_\pi} \rp)^2 
\vec{\tau}_1\vec{\tau}_2 \bm{\Sigma}_1 \cdot \bm{\Sigma}_2 
\delta (\bm{r}_1-\bm{r}_2)\; ,
\label{deltapi}
\end{equation} 
with the parameter $g^{\prime}$ adjusted in such a way that the PN-RQRPA
reproduces the experimental values of GTR excitation energies~\cite{PNVR.04}.
For the DD-ME2 effective interaction the value of  $g^{\prime}$=0.52 has 
been adjusted to the position of the GTR in $^{208}$Pb.

In most applications of the RHB and RQRPA models \cite{Vre.05,PVKC.07},
the pairing part of the Gogny force has been used in the 
the $pp$-channel:
\begin{equation}
V^{pp}(1,2)~=~\sum_{i=1,2}e^{-[(\mathbf{r}_{1}-\mathbf{r}_{2})/{\mu _{i}}%
]^{2}}\,(W_{i}~+~B_{i}P^{\sigma }-H_{i}P^{\tau }-M_{i}P^{\sigma }P^{\tau }),
\label{Gogny}
\end{equation}
with the set D1S \cite{BGG.91} for the parameters $\mu _{i}$, $W_{i}$, $%
B_{i} $, $H_{i}$ and $M_{i}$ $(i=1,2)$. This force has been very carefully
adjusted to pairing properties of finite nuclei all over the periodic
table. In particular, the basic advantage of the Gogny force is the finite
range, which automatically guarantees a proper cut-off in momentum space.

%----------------------------------------------------------------------------
\section{\label{sectionIV}
Illustrative calculations: cross sections for neutrino detector response}
%----------------------------------------------------------------------------

The theoretical framework described in the previous two sections can 
systematically be applied in studies of charged-current neutrino reaction 
rates with target nuclei of arbitrary mass. We have performed 
several illustrative calculations of cross sections for reactions with 
nuclei of interest for neutrino detector response: $^{12}$C, $^{16}$O, 
$^{56}$Fe, and $^{208}$Pb.  The inclusive cross sections, 
which sum the contributions from transitions to all possible final states, 
are given as functions of neutrino energy, and can be averaged 
over the experimental neutrino flux when available, e.g. from the 
decay at rest of $\mu^+$, and the decay in flight of 
$\pi^+$~\cite{Ath.97,Bod.92,Bod.94,Mas.98,Kra.92,Koe.92}.
The flux-averaged cross sections provide a crucial test for the
validity of the theoretical approach used for modeling 
neutrino-nucleus reactions.

For charged-current reactions the cross section Eq.~(\ref{CS}) must be corrected 
for the distortion of the outgoing lepton wave 
function by the Coulomb field of the daughter nucleus. 
In order to be able to compare our results with previous studies \cite{Vol.00,Kol.03}, 
we use the same prescription for the Coulomb correction. The cross section can either 
be multiplied by a Fermi function obtained from the numerical solution of the Dirac 
equation for an extended nuclear charge distribution \cite{Kol.03} or, at higher energies, 
the effect of the Coulomb field can be described by the effective momentum approximation (EMA) \cite{Eng.98,Tra.01,Ast.04}, in which the lepton momentum 
$p_l$ and energy $E_l$ are modified by a constant electrostatic potential evaluated 
at the center of the nucleus:
\begin{equation}
p_l^{eff} = \sqrt{E_l^{eff}-m_l^2}
\label{peff}
\end{equation}
\begin{equation}
E_l^{eff} = E_l - V_C(0),
\label{eeff} 
\end{equation}
where $V_C(0)=- { 3Z\alpha}/{(2r_c) }$, the nuclear charge radius is 
denoted by $r_c$, and $m_l$ is the mass of the outgoing lepton.
In calculations with EMA the original lepton momentum $p_l$ and 
energy $E_l$ appearing in the expression for
the cross section, are replaced by the above effective quantities.

\begin{figure}
%\vspace*{1cm}
\centering
\includegraphics[scale=0.5]{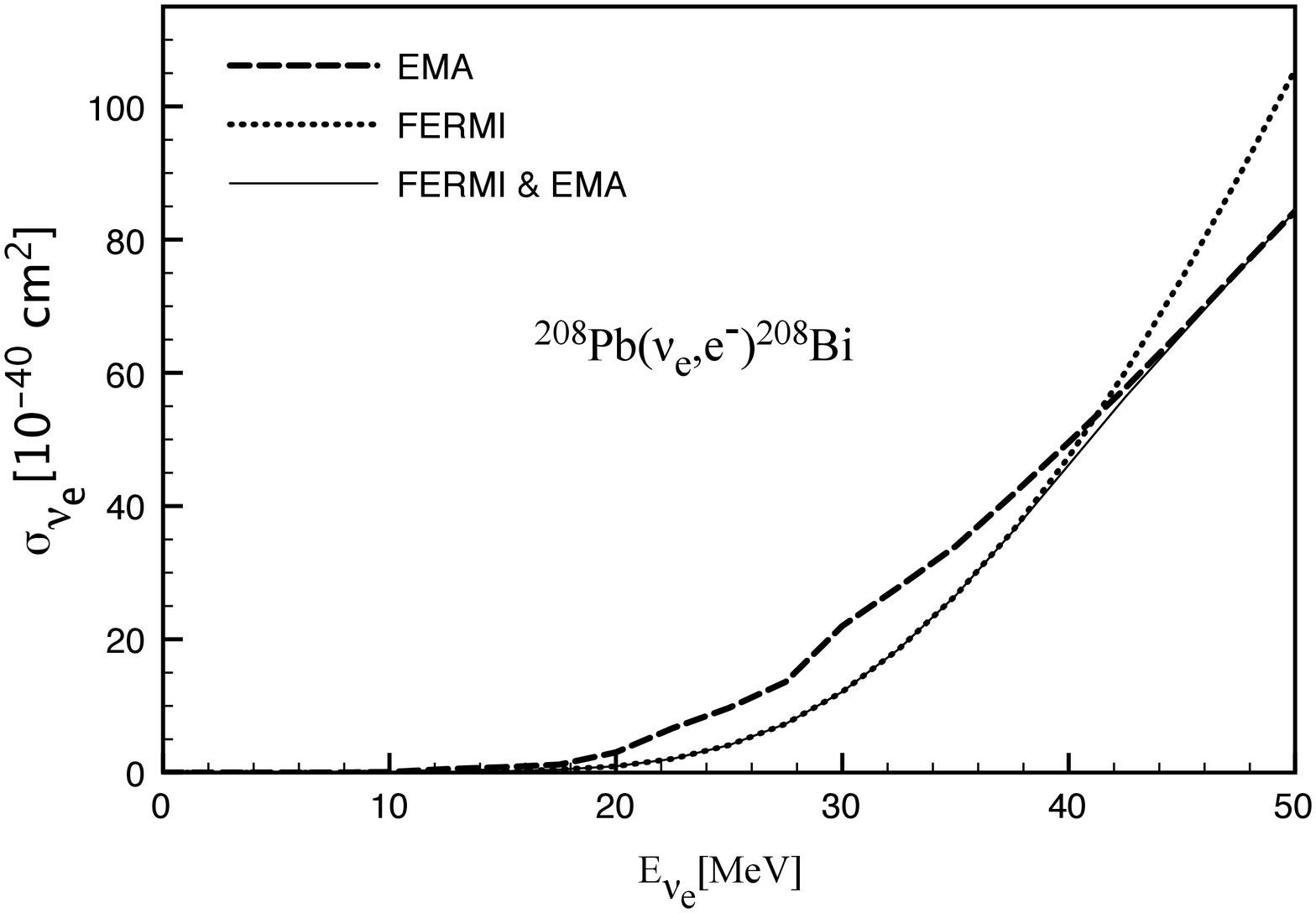}
\caption{Inclusive neutrino-nucleus cross sections for the 
$^{208}$Pb($\nu_e,e^-$)$^{208}$Bi reaction, evaluated using the 
effective momentum approximation (dashed), the Fermi function 
correction (dotted), and their combination (solid).
The PN-RQRPA calculation is performed with the DD-ME2 
effective interaction.}
\label{csm_Pb208}
\end{figure}

The two prescriptions for the Coulomb correction are tested in the calculation 
of the inclusive cross section for neutrino scattering on a high-Z nucleus:
$^{208}$Pb($\nu_e,e^-$)$^{208}$Bi.
In Fig.~\ref{csm_Pb208} we display the corresponding cross sections corrected 
either with the Fermi function, or calculated with the EMA. As already shown in 
previous studies, the Fermi function correction overestimates the cross sections at 
higher neutrino energies, where the EMA provides a more reliable correction. 
Therefore, following the prescription from Ref.~\cite{Vol.00}, we will use the Fermi 
function correction in the range of neutrino energies for which the cross section is 
below the corresponding EMA value, whereas the EMA will 
be employed at higher energies (cf. the solid curve in Fig.~\ref{csm_Pb208}). 

\begin{figure}
%\vspace*{1cm}
\centering
\includegraphics[scale=0.5]{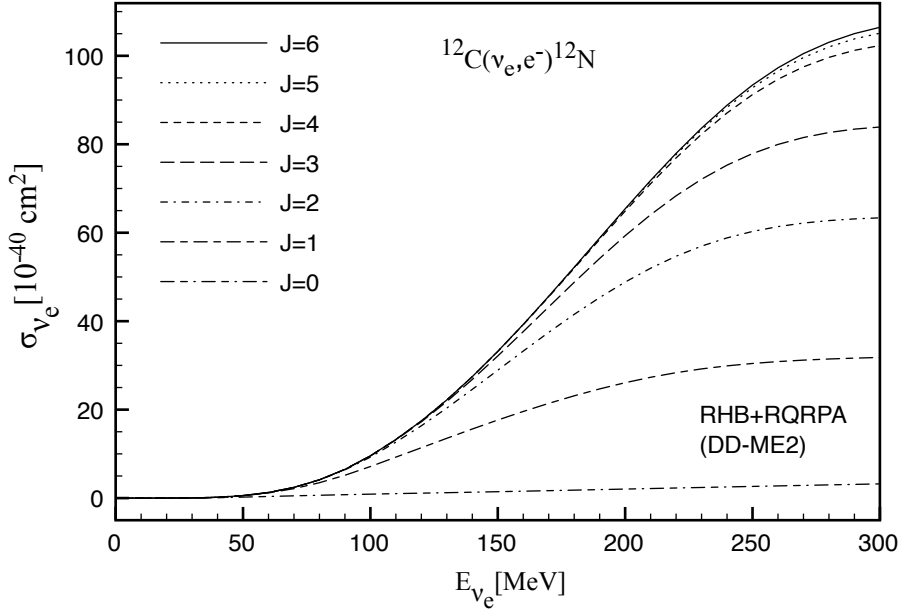}
\caption{The RHB plus PN-RQRPA inclusive neutrino-nucleus cross sections for the 
$^{12}$C($\nu_e,e^-$)$^{12}$N reaction.  The different curves correspond 
to cross sections evaluated by successively increasing the 
maximal allowed angular momentum in the sum over $J$ in Eq.~(\ref{ncs_eq}):
from $J_{max}=0^\pm$ to $J_{max}= 7^\pm$. }
\label{csm_C12_Jmax7}
\end{figure}

One of the most extensively studied neutrino-nucleus reactions is 
$^{12}$C($\nu_e,e^-$)$^{12}$N. This reaction is particularly important 
because $^{12}$C is used in liquid scintillator detectors, and data on the cross section are available from the LSND~\cite{Alb.95,Ath.97}, and 
KARMEN~\cite{Bod.92,Bod.94,Mas.98} collaborations, as well as from 
LAMPF~\cite{Kra.92,Koe.92}. In order to illustrate the contributions of 
different multipole excitations, in Fig.~\ref{csm_C12_Jmax7} we plot 
the inclusive cross section for the $^{12}$C($\nu_e,e^-$)$^{12}$N reaction 
as function of the neutrino energy, obtained by successively increasing the 
maximal allowed angular momentum in the sum over $J$ in Eq.~(\ref{ncs_eq}):
from $J_{max}=0^\pm$ to $J_{max}= 7^\pm$. One notices that the largest 
contributions arise from $J=1^\pm$ and $J= 2^\pm$, and that the 
contribution of higher multipolarities gradually decreases. In fact, in this 
figure one can not distinguish the cross sections calculated with 
$J_{max}= 6^\pm$ and $J_{max}= 7^\pm$, for the whole interval of 
neutrino energies.
 
The results of model calculations can be compared with available data 
by averaging the cross section over the neutrino flux $f(E_{\nu})$, 
which depends on the specific neutrino source
\begin{equation}
\langle \sigma_{\nu} \rangle =  \frac{ \int dE_{\nu}
\sigma_{\nu}(E_{\nu})f(E_{\nu})  }{\int dE'_{\nu}f(E'_{\nu}) } \;.
\end{equation}
For $\nu_e$ the Michel flux from the decay at rest~(DAR) of $\mu^+$
is used \cite{Kra.92},
\begin{equation}
f(E_{\nu_e}) =  \frac{96 E_{\nu_e}^2}{m_{\mu}^4}  \left (  m_{\mu} -2
E_{\nu_e}  \right ) \;,
\label{darflux}
\end{equation}
whereas for $\nu_{\mu}$ we employ the polynomial fit to the experimental
flux obtained from the decay in flight~(DIF) of $\pi^+$~\cite{LSND.98}.
In Table~\ref{table1} the values of the muon and electron neutrino-nucleus
cross sections for the reactions $^{12}$C($\nu_e,e^-$)$^{12}$N and
$^{12}$C($\nu_{\mu},{\mu}^-$)$^{12}$N, calculated with the PN-RQRPA 
and averaged over the empirical neutrino fluxes, are compared with the
results of previous theoretical studies based on the
non-relativistic random phase approximation and the shell model,  
and with the experimental values:  LAMPF~\cite{Kra.92}, 
KARMEN~\cite{Bod.94} and LSND~\cite{Ath.97,LSND.98,LSND.01}.
For the ($\nu_{\mu},{\mu}^-$) reaction the 
PN-RQRPA cross section is in excellent agreement with the 
(Q)RPA results of Ref.~\cite{Vol.00}, and the CRPA value of 
Ref.~\cite{Kol.99}, but somewhat larger than the experimental cross section.
The PN-RQRPA cross section for the $^{12}$C($\nu_e,e^-$)$^{12}$N 
reaction is in better agreement with experiment when compared 
with the results of the QRPA calculations~\cite{Vol.00,Krm.05}. We notice 
that a very good agreement with data has been obtained in the 
shell-model calculations of Refs.~\cite{Hay.00,Vol.00,Aue.02}, and
with the RPA based on the Skyrme interactions SGII and SIII \cite{Aue.97}.

\begin{figure}
%\vspace*{1cm}
\centering
\includegraphics[scale=0.5]{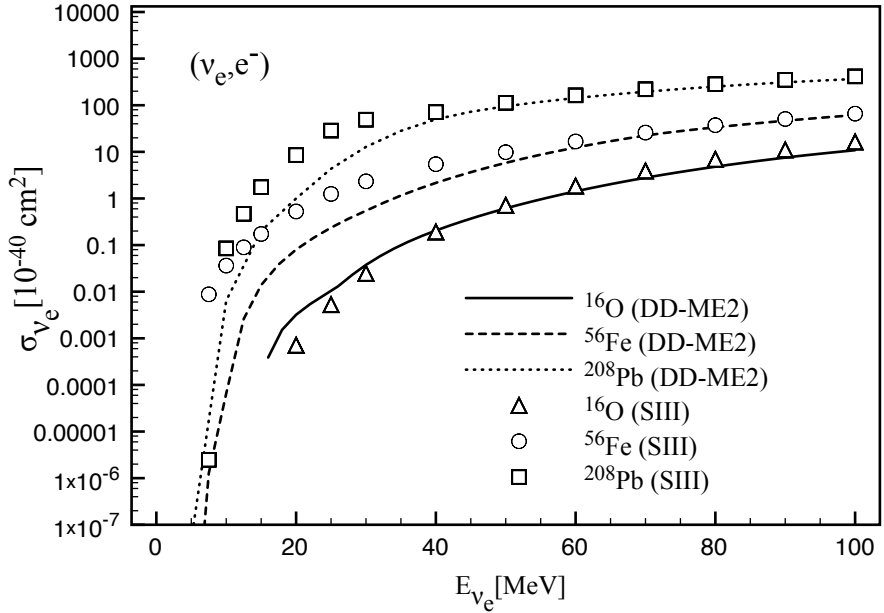}
\caption{Inclusive cross sections for the $^{16}$O($\nu_e,e^-$)$^{16}$F,  
$^{40}$Ar($\nu_e,e^-$)$^{40}$K,
and $^{208}$Pb($\nu_e,e^-$)$^{208}$Bi reactions. 
The RHB plus PN-RQRPA(DD-ME2) results are compared with those
obtained in the QRPA with the
Skyrme interaction SIII~\cite{Laz.07}.}
\label{csm_all}
\end{figure}

In Fig.~\ref{csm_all} the calculated inclusive cross sections are plotted up to 
100 MeV neutrino energy for three reactions including also heavier nuclei:
$^{16}$O($\nu_e,e^-$)$^{16}$F, $^{56}$Fe($\nu_e,e^-$)$^{56}$Co,
and $^{208}$Pb($\nu_e,e^-$)$^{208}$Bi. The results of the RHB 
plus PN-RQRPA calculations
are compared with those of a very recent analysis performed with 
the nonrelativistic QRPA and using the Skyrme interaction SIII \cite{Laz.07}.
First we notice the pronounced enhancement of cross sections for 
neutrino reactions on heavier nuclei: from $^{16}$O to $^{208}$Pb
the reaction cross sections increase by more than two orders of 
magnitude. For the reaction on $^{16}$O the present results are
in excellent agreement with the QRPA - SIII cross sections \cite{Laz.07} 
at all neutrino energies. This is also the case for the reactions on
$^{56}$Fe and $^{208}$Pb at neutrino energies above 40 MeV, whereas 
at lower energies the RHB plus RQRPA cross sections appear somewhat
below the QRPA results. 

In table~\ref{table2} we compare various theoretical results of 
flux-averaged cross sections for the electron neutrino reactions 
with $^{16}$O, $^{56}$Fe, and $^{208}$Pb target nuclei. 
For the neutrino flux the DAR spectrum Eq.~(\ref{darflux}) is
used, and values obtained with the shell model \cite{Aue.02}, 
the tensor model \cite{Min.02}, Tamm-Dancoff \cite{Suz.03},  
and various RPA-based calculations are included.
For the $^{16}$O target, the present PN-RQRPA(DD-ME2)
cross section agrees with the shell model and (Q)RPA results.
For heavier targets the differences between various 
calculations are more pronounced. Particularly important 
is the reaction  $^{56}$Fe($\nu_e,e^-$)$^{56}$Co, because it is
the only case in which data are available for a medium-heavy target 
nucleus \cite{Mas.98,Kol.01}. The two self-consistent  
calculations: the QRPA~\cite{Laz.07} and the present 
RHB plus PN-RQRPA, which do not require
any additional adjustments of the model parameters to the  
specific target nucleus, predict values for the cross section: 
352$\times$10$^{-42}$cm$^2$(QRPA~\cite{Laz.07}) and 
140$\times$10$^{-42}$cm$^2$(PN-RQRPA) which differ 
considerably, but are still within the uncertainties of the 
experimental result (256$\pm$108$\pm$43)$\times$10$^{-42}$cm$^2$.

Obviously various theoretical approaches differ significantly
in the predicted neutrino-nucleus cross sections, and this will 
require more detailed studies of the underlying nuclear structure that 
contributes to the neutrino reaction rates. On the other hand, 
the only data for a medium-heavy target nucleus are from 
the KARMEN collaboration~\cite{Mas.98}, and this result has not 
yet been confirmed by independent measurement. One hopes 
that future experiments will provide additional constraints for 
theoretical models that are used in the description of weak interaction 
rates. At the same time the nuclear structure input for 
neutrino-nucleus cross sections must be further analyzed, and
checked by improving the description of
nuclear ground-state properties and spin-isospin excitations 
relevant for the calculation of the reaction rates.

\begin{figure}
%\vspace*{1cm}
\centering
\includegraphics[scale=0.5]{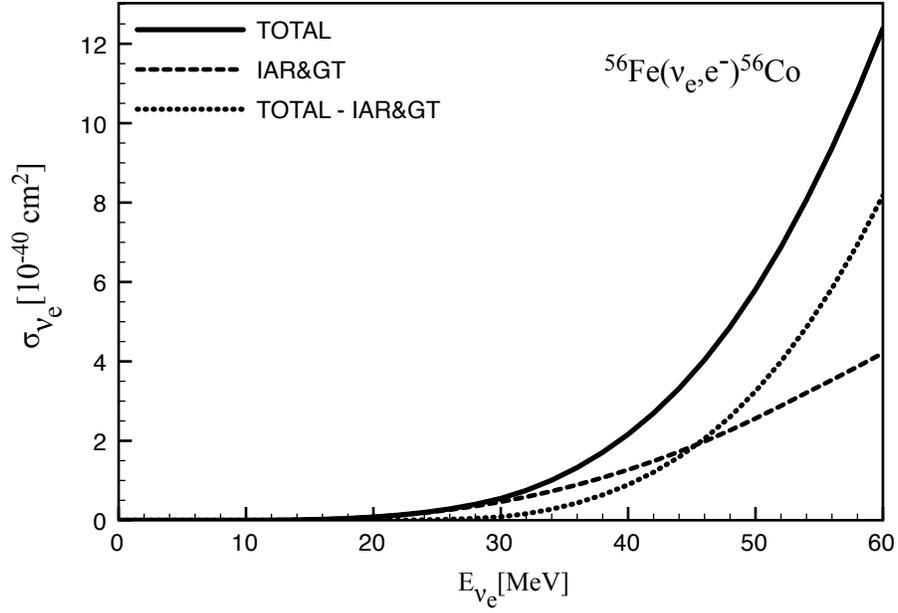}
%\vspace*{0.5cm}
\caption{Inclusive cross section for the $^{56}$Fe($\nu_e,e^-$)$^{56}$Co reaction. 
The solid curve corresponds to the full calculation 
including all transitions with multipolarity  $J \leq 7^\pm$, whereas 
the cross section evaluated for the IAS and GT transitions only   
is represented by the dashed curve. The dotted curve is the 
difference between the two calculations.}
\label{csm_Fe56}
\end{figure}

Next we consider in more detail the 
contributions from excitations of various multipolarities
to the neutrino-nucleus cross sections evaluated within the RHB 
plus PN-RQRPA framework. Fig.~\ref{csm_Fe56} displays the 
calculated cross sections for the
reaction $^{56}$Fe($\nu_e,e^-$)$^{56}$Co. The solid curve 
represents the inclusive cross section as a function of the 
neutrino energy calculated by including  
all multipole transitions with $J \leq 7^\pm$, whereas 
the dashed curve corresponds to the
cross section evaluated only for transitions to the
isobaric analog state (IAS) and Gamow-Teller (GT) resonance. 
The difference between these two cases is represented 
by the dotted curve which, as one notes, rises sharply with energy. 
Below an incident neutrino energy of
$E_{\nu_e}\approx$ 30 MeV the dominant contribution to the cross section 
originates from the two basic charge exchange modes: IAS and GTR. 
With increasing $E_{\nu_e}$, however, contributions from other multipole 
transitions become even more important. This result clearly shows that
models which take into account only the IAS and GTR transitions~\cite{Bor.00}, 
cannot provide quantitative predictions of cross sections at neutrino 
energies $E_{\nu_e} \geq $ 30 MeV.

\begin{figure}
 % \hfill
\begin{center}
  \begin{minipage}[t]{.45\textwidth}
   \begin{center} 
      \includegraphics[scale=0.3]{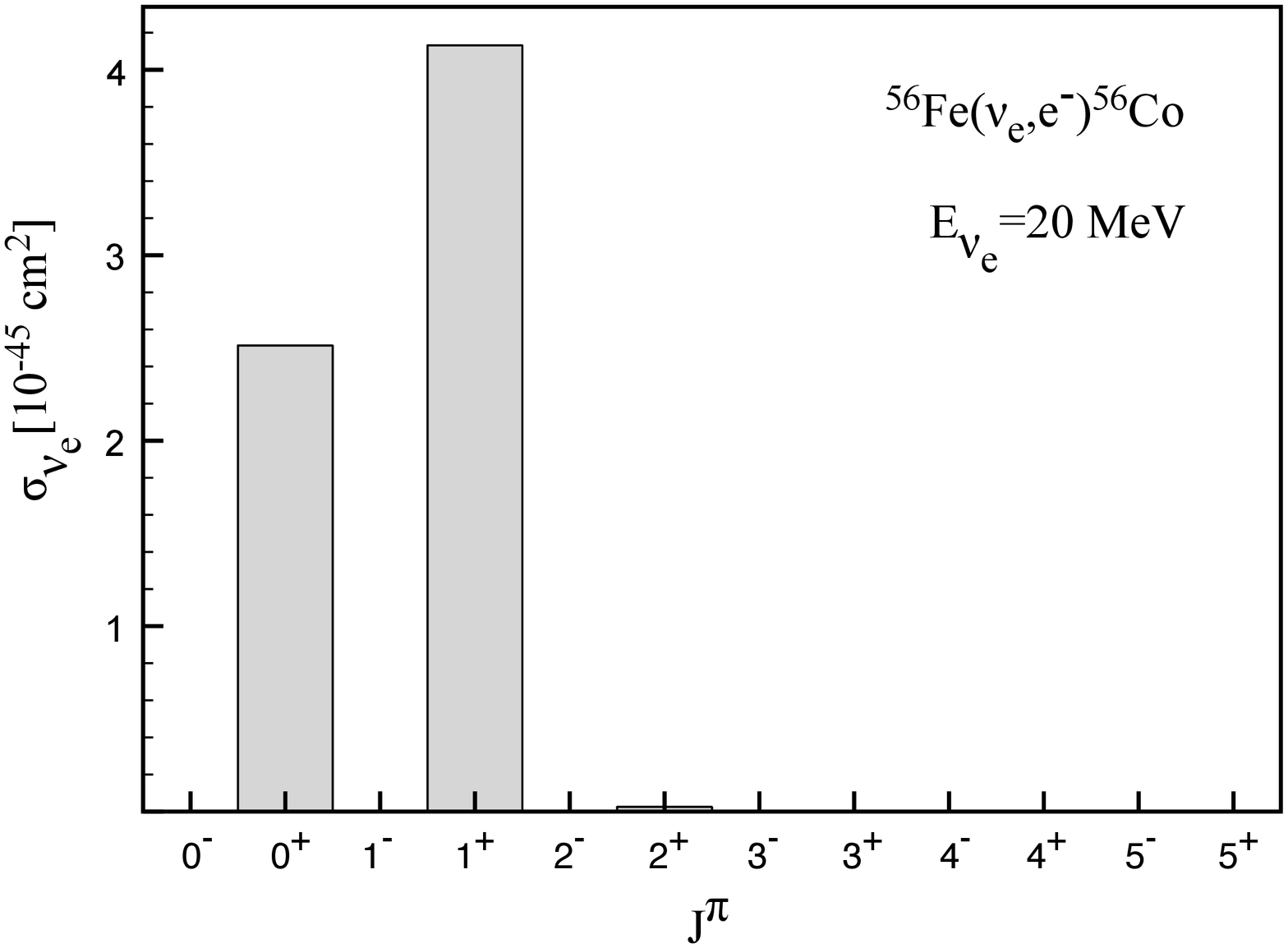}
    \end{center}
  \end{minipage}
 % \hfill
  \begin{minipage}[t]{.45\textwidth}
    \begin{center}
     \includegraphics[scale=0.3]{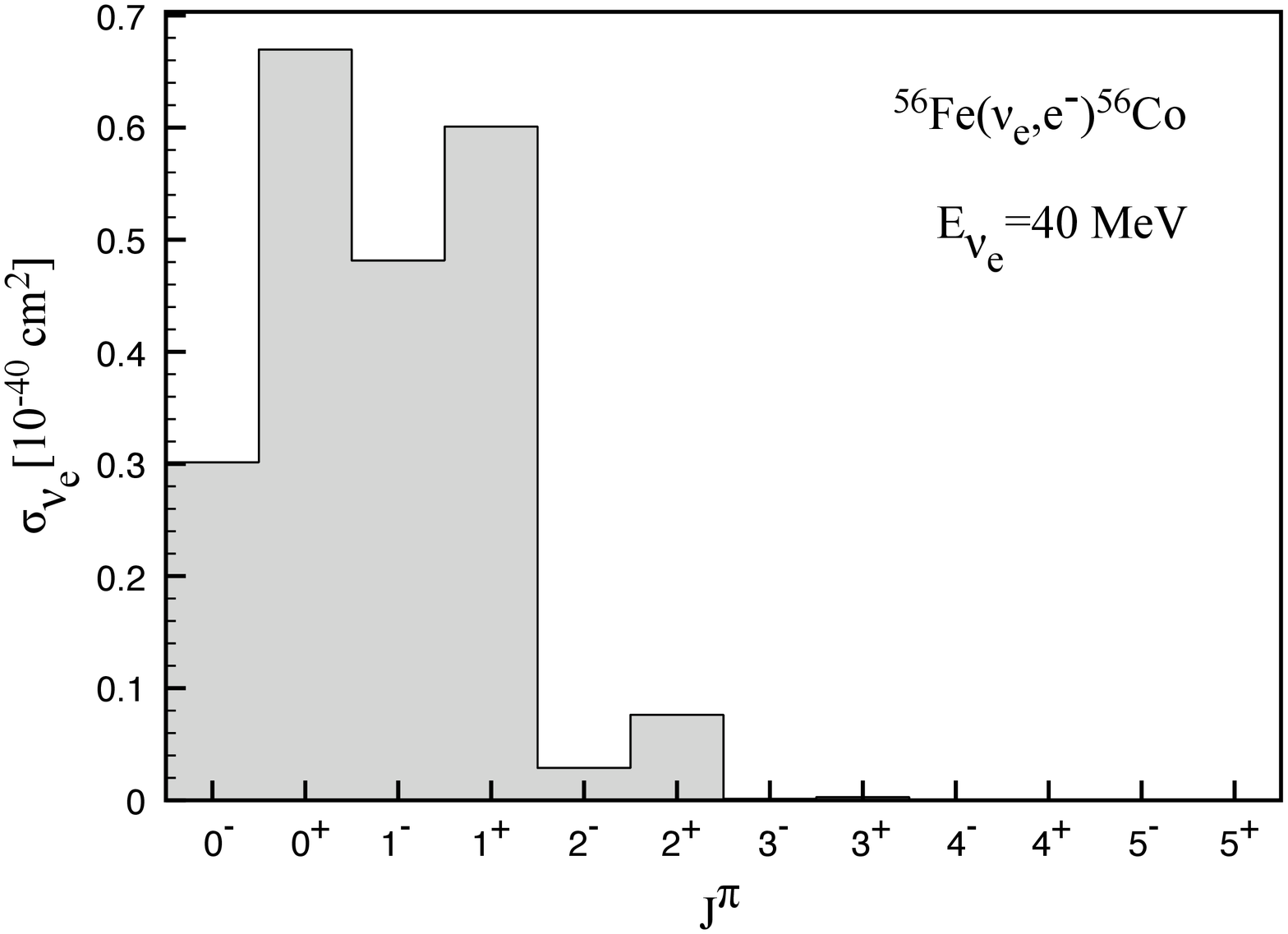}
    \end{center}
  \end{minipage}
  \begin{minipage}[t]{.45\textwidth}
    \begin{center}  
      \includegraphics[scale=0.3]{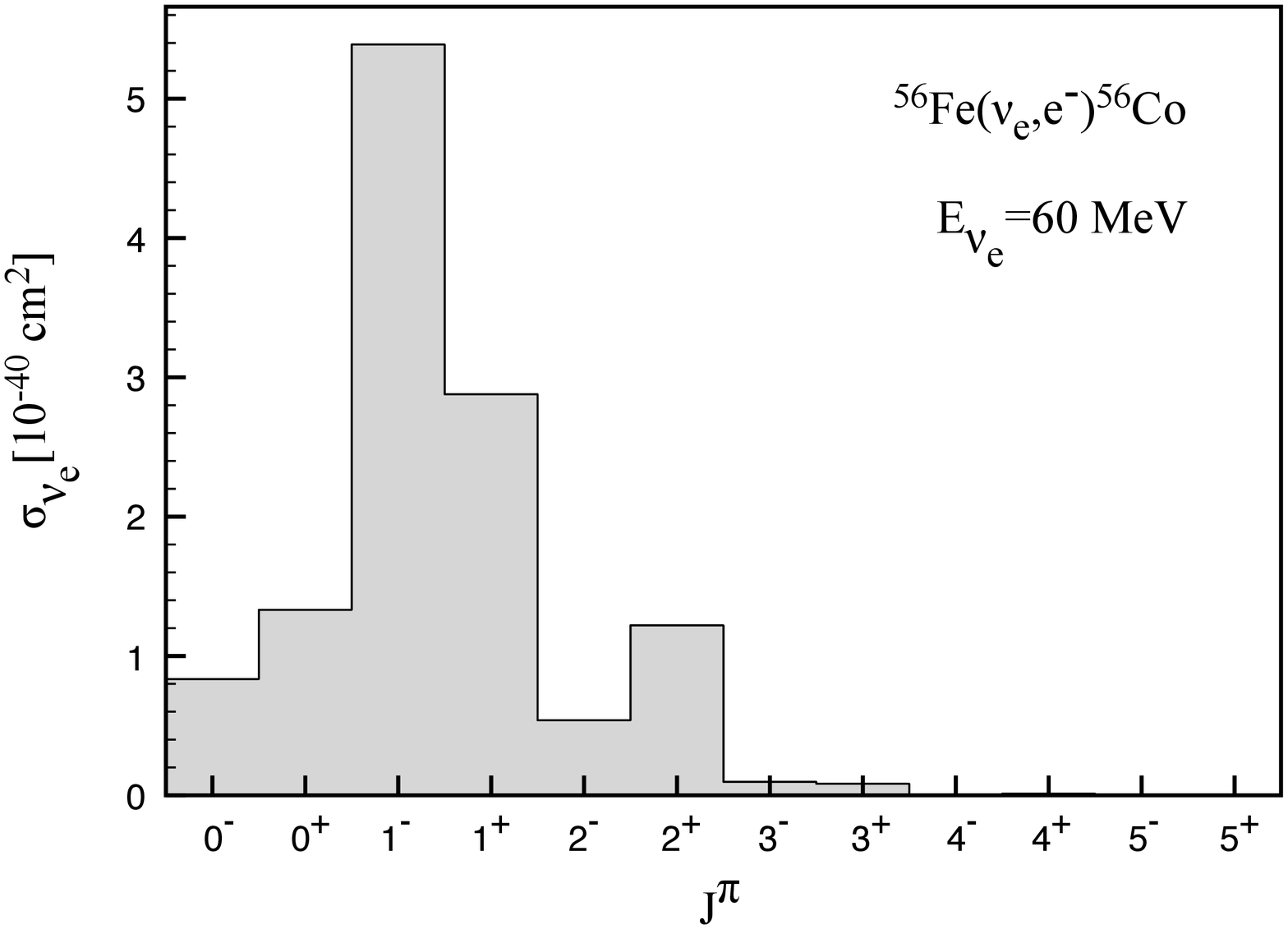}
    \end{center}
  \end{minipage}
  %\hfill
  \begin{minipage}[t]{.45\textwidth}
    \begin{center}  
      \includegraphics[scale=0.3]{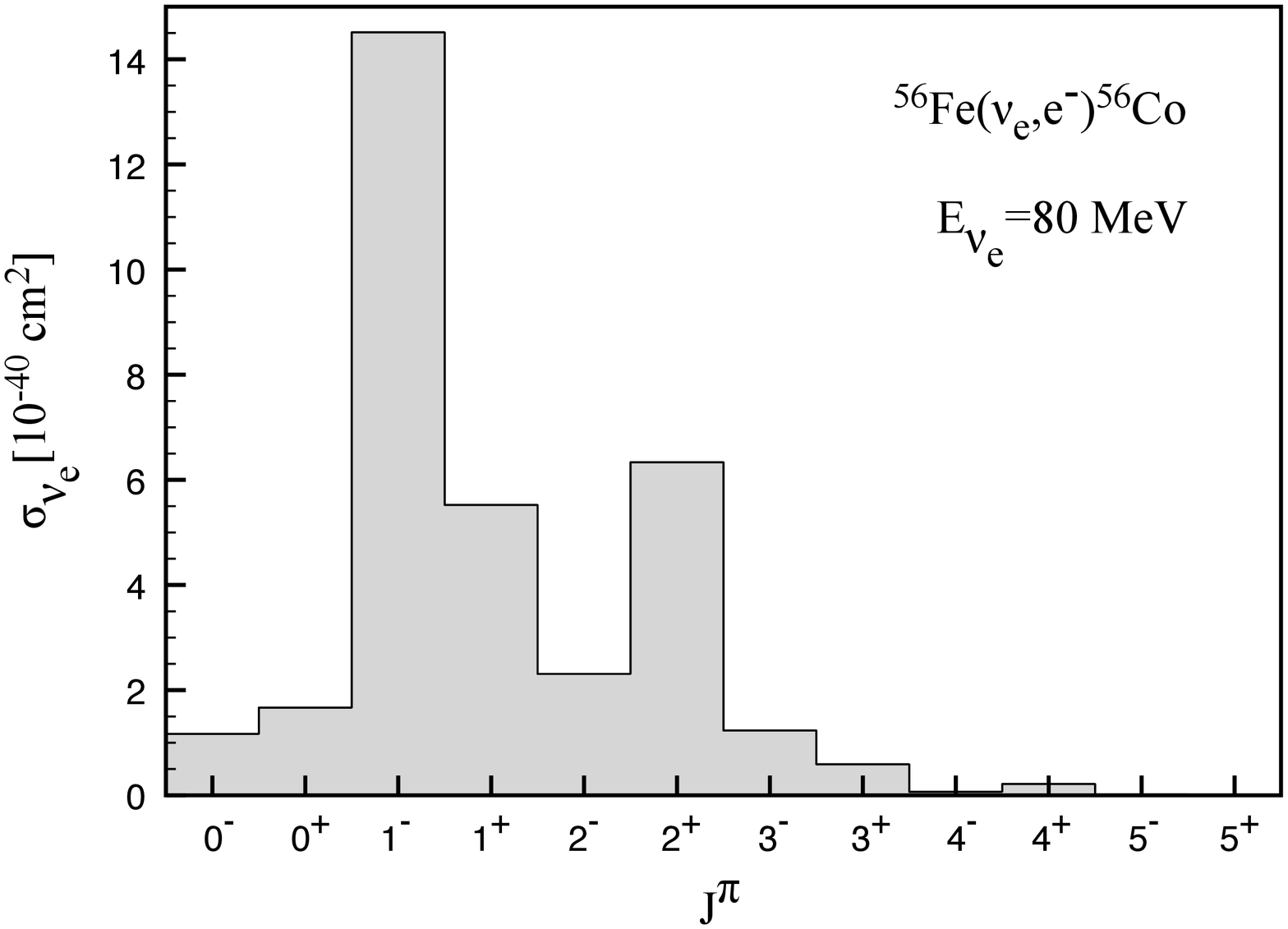}
    \end{center}
  \end{minipage}

  \caption{\label{Fe56distr}Contributions of different multipole transitions to the RHB 
  plus PN-RQRPA inclusive cross section for the $^{56}$Fe($\nu_e,e^-$)$^{56}$Co reaction, at the incoming electron 
neutrino energies $E_{\nu_e}$ = 20, 40, 60, and 80 MeV.}
  \hfill
\end{center}
%\label{Fe56distr}
\end{figure}

The contribution of different multipole transitions to the cross section for 
the reaction $^{56}$Fe($\nu_e,e^-$)$^{56}$Co, at increasing neutrino 
energies, is illustrated in Fig.~\ref{Fe56distr}. For the neutrino energies 
$E_{\nu_e}$ = 20, 40, 60, and 80 MeV, we display 
the partial contributions to the total cross section of multipoles from $J = 0^\pm$ to 
$J = 5^\pm$. At $E_{\nu_e} = 20$ MeV the reaction is dominated by transitions to the
IAS and GTR. At higher energies, however, other multipolarities 
start to play an important role and one notices, in particular, the dominant 
contribution of the $1^-$ transitions at $E_{\nu_e}$=60 and 80 MeV. 
This distribution of multipole transitions, calculated in 
the RHB plus PN-RQRPA  model, is similar to the one obtained 
in the QRPA with Skyrme effective interactions~\cite{Laz.07}. 
The differences in the contributions of higher multipoles can be 
attributed to the different effective interactions employed 
in the two models.

%----------------------------------------------------------------------------
\section{\label{sectionV}Cross sections for supernova neutrinos}
%----------------------------------------------------------------------------

An important application of microscopic models of neutrino-nucleus 
reactions is the calculation of cross sections for supernova neutrinos.
Accurate modeling of reaction rates on nuclei that can be used as 
targets for the supernova neutrino detectors is, of course, essential 
for studies of supernova dynamics and, in particular, of weak 
interaction processes  which determine the evolution of a
supernova explosion. In this section we compare the results 
of illustrative RHB plus PN-RQRPA calculations of charged-current 
reaction rates with those obtained in the shell model~\cite{Hax.87}, 
and with the CRPA~\cite{Kol.03}. The supernova neutrino flux is 
usually described by the Fermi-Dirac spectrum:
\begin{equation}
f(E_{\nu}) =  \frac{1}{T^3}  \frac{E_{\nu}^2}
{exp \left[ (E_{\nu}/T)-\alpha  \right ] + 1  } \; .
\label{fermidirac}
\end{equation}

\begin{figure}
%\vspace*{1cm}
\centering
\includegraphics[scale=0.5]{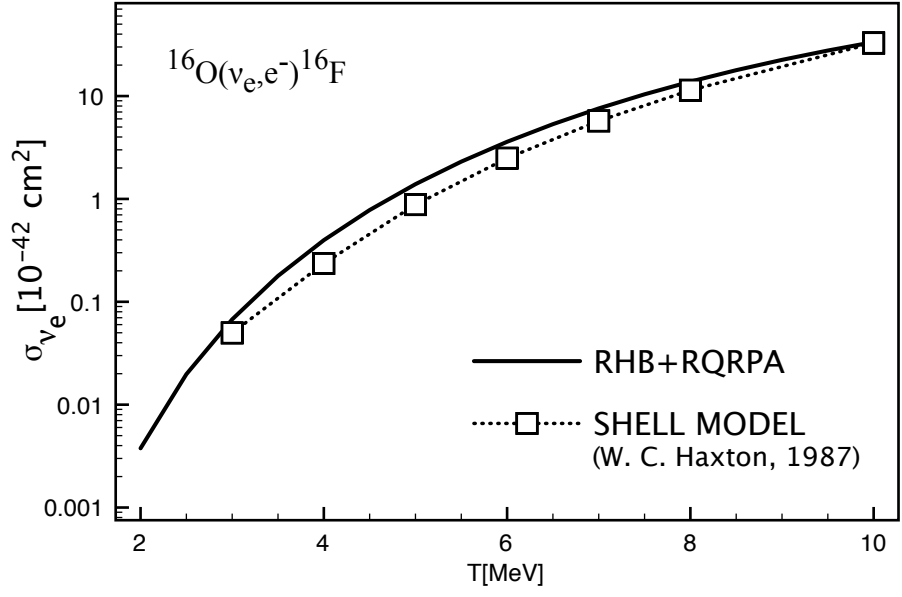}
%\vspace*{0.5cm}
\caption{Inclusive cross section for the $^{16}$O($\nu_e,e^-$)$^{16}$F reaction, averaged over the supernova neutrino flux and plotted as a function of temperature. 
The PN-RRPA results obtained with the DD-ME2 effective interaction are compared 
with the shell-model cross-sections from Ref.~\cite{Hax.87}.}
\label{O16_Temp}
\end{figure}

At typical supernova neutrino energies one expects the total 
cross section for the charged current reaction ($\nu_e,e^-$) 
to be dominated by the allowed transitions to the IAS and 
the Gamow-Teller resonance states in the daughter nucleus.
In Fig.~\ref{O16_Temp} we display the 
flux-averaged cross sections for the reaction 
$^{16}$O($\nu_e,e^-$)$^{16}$F, evaluated at
different temperatures in the interval $T= 2 - 10$ MeV, and for
the chemical potential $\alpha$=0. The results
are in a remarkable agreement with those obtained in the shell-model
calculation of Ref.~\cite{Hax.87}. This is a very interesting result, 
because the two models are based on entirely different concepts, and use 
different effective interactions. An important advantage of the present approach, 
however, is that it can easily be extended to heavier nuclei and to systems 
far from stability, and therefore also applied in the description of the 
$r$-process nucleosynthesis. 

\begin{figure}
%\vspace*{1cm}
\centering
\includegraphics[scale=0.5]{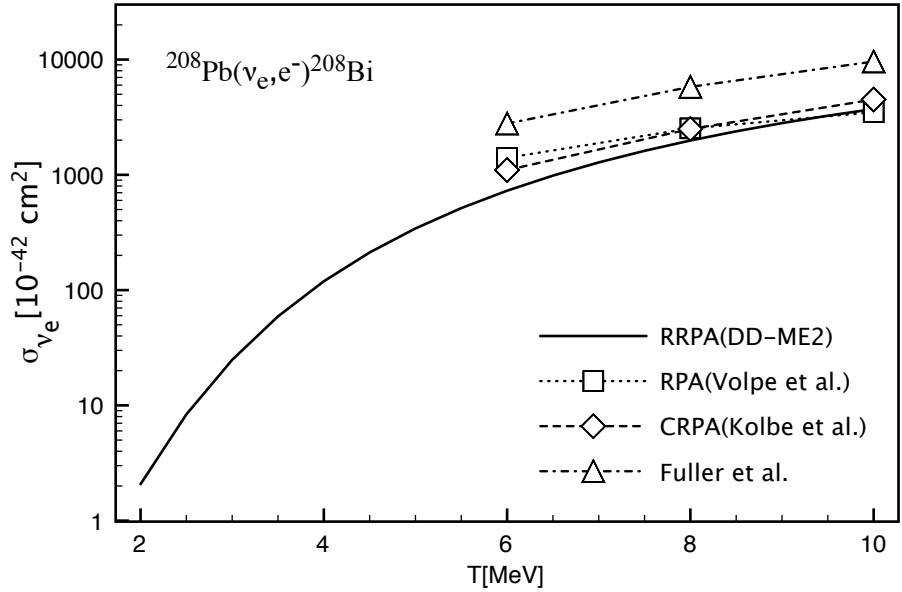}
%\vspace*{0.5cm}
\caption{Same as in Fig.~\ref{O16_Temp}, but for the 
$^{208}$Pb($\nu_e,e^-$)$^{208}$Bi reaction, in comparison with
the results from Refs.~\cite{Vol.00,Kol.03,Ful.99}.}
\label{Pb208_Temp}
\end{figure}

We have further tested the model in the case of the heavy target: 
$^{208}$Pb($\nu_e,e^-$)$^{208}$Bi.
For the same neutrino flux as in the previous example, the 
corresponding cross sections are shown in Fig.~\ref{Pb208_Temp}
and compared, at $T=6,8$ and 10 MeV, with the results from 
Refs.~\cite{Vol.00,Kol.03,Ful.99}. Since at higher temperatures the 
neutrino flux extends toward higher neutrino energies, 
the calculated cross sections display a pronounced enhancement
with temperature (cf. Sec.~\ref{sectionIV}).
In comparison with the RPA-based results from
Refs.~\cite{Vol.00,Kol.03}, the present
PN-RQRPA calculation predicts a slightly lower value for the
flux-averaged cross section at $T=6$ MeV, whereas an 
excellent agreement is obtained at 
$T=8$ and 10 MeV. Note that the calculation of Ref.~\cite{Ful.99},
which includes the IAS, the GT, and the first-forbidden transitions 
computed in the Goldhaber-Teller model, consistently predicts higher
values for the flux-averaged cross section at all considered temperatures.
As it has already been discussed in Ref.~\cite{Vol.02}, a possible origin of 
this difference could be the use of the Fermi function for the Coulomb
correction at higher neutrino energies, instead of the EMA.

\begin{figure}
%\vspace*{1cm}
\centering
\includegraphics{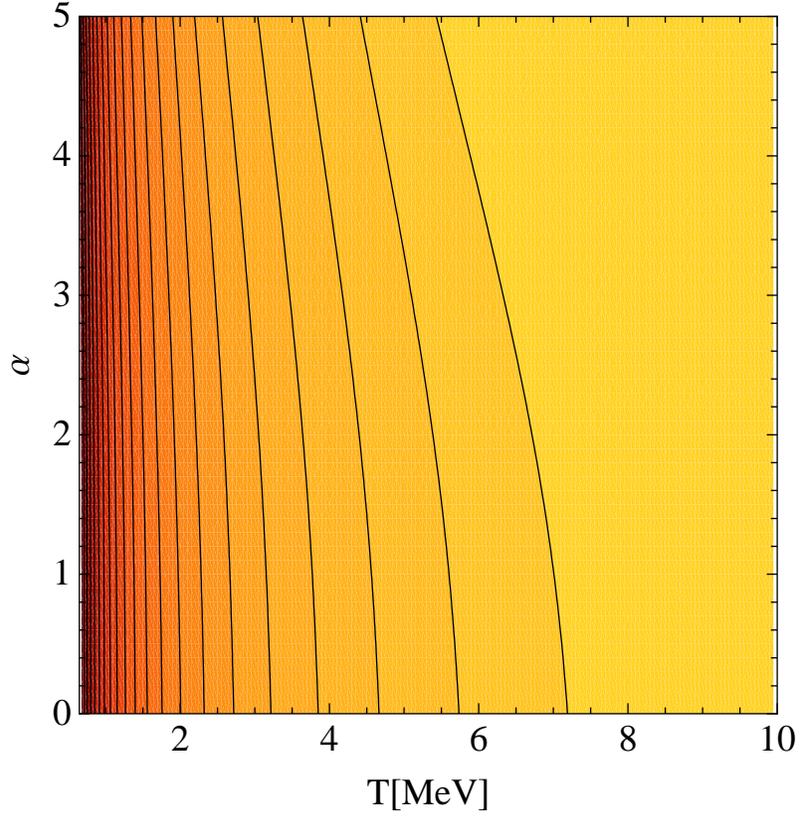}
%\vspace*{0.5cm}
\caption{Contour plot of supernova neutrino-nucleus cross sections 
for the $^{16}$O($\nu_e,e^-$)$^{16}$F reaction, as functions of 
the temperature T and chemical potential $\alpha$ which determine 
the Fermi-Dirac neutrino spectrum Eq.~(\ref{fermidirac}).
The solid curves in the plot correspond to constant values of the
cross section, and lighter shading denotes the increase
of cross sections.}
\label{O16_contour}
\end{figure}

Although most calculations of neutrino-nucleus
cross sections consider only the case when the 
chemical potential $\alpha$=0, in astrophysical 
applications it might be important to perform 
studies of reaction rates for other values 
of $\alpha$ \cite{Kol.03}. We have therefore explored the 
sensitivity of the neutrino reaction cross sections to $\alpha$. 
In Fig.~\ref{O16_contour} the contour plot is shown for the 
$^{16}$O($\nu_e,e^-$)$^{16}$F cross sections as
functions of $T$ and $\alpha$ in the Fermi-Dirac spectrum
of Eq.~(\ref{fermidirac}). The curves in the plot correspond 
to constant values of the cross sections. At lower temperatures
the cross sections depend only very weakly on $\alpha$. However,
already above T=2 MeV, the cross sections increase much faster
for higher values of the chemical potential $\alpha$.

%----------------------------------------------------------------------------
\section{\label{sectionVI}Concluding remarks}
%----------------------------------------------------------------------------
Detailed microscopic calculations of charged-current and 
neutral-current neutrino-nucleus reaction rates are of crucial importance  
for models of neutrino oscillations, detection of supernova neutrinos, 
and studies of the r-process nucleosynthesis. In this work we have 
introduced a consistent microscopic method for calculating 
neutrino-nucleus cross sections, based on the relativistic 
mean-field framework. In this approach the ground state 
of a nucleus is described with the relativistic Hartree-Bogoliubov 
model, and the neutrino-induced transitions to 
excited nuclear states are computed in the relativistic quasiparticle 
random phase approximation. Since it is based on the 
self-consistent mean-field approach to nuclear structure, 
the model can be applied to neutrino reactions with 
target nuclei of arbitrary mass throughout the chart of nuclides. 
By employing universal effective interactions, with  
parameters adjusted to global nuclear properties, the calculation 
of neutrino-nucleus cross sections is essentially parameter free. 
In the particular model introduced in this work, once the  
ground state of a target nucleus is calculated with the RHB model, 
the only additional parameter in the PN-QRPA calculations of 
excited states is the strength $g^{\prime}$ of the zero-range
Landau-Migdal term, which accounts for the contact part of the 
nucleon-nucleon interaction. The value of this parameter 
depends on the isovector properties (symmetry energy) of the 
effective interaction in the particle-hole channel, and is adjusted 
to reproduce the excitation energy of the GTR in $^{208}$Pb. 
The same value of $g^{\prime}$ is then used in calculations of charge-exchange 
excitations in all other nuclei. Since it does not contain parameters 
adjusted to specific target nuclei, or even to a particular mass region, 
the model can be extended to regions of nuclei far from stability for 
which no data on ground and/or excited states are available, including 
those on the path of the r-process.

We have performed a number of illustrative test calculations for
charged-current neutrino reactions  on $^{12}$C, $^{16}$O, 
$^{56}$Fe, and $^{208}$Pb, in the low-energy range below 100 MeV 
neutrino energy. The results have been compared with those 
obtained in previous theoretical studies based on the shell model 
and the non-relativistic random phase approximation,  
and with the available experimental values for flux-averaged 
cross sections. In addition to the total neutrino-nucleus cross 
sections, we have also analyzed the evolution of the contributions 
of different multipole excitations as a function of neutrino energy.  
It has been shown that except at relatively low neutrino energies 
$E\leq 30$ MeV, for which the reactions are dominated by transitions 
to IAS and GTR states, at higher energies the inclusion of spin-dipole 
transitions, and also excitations of higher multipolarities, is essential for 
a quantitative description of neutrino-nucleus cross sections. Finally, 
we have also investigated the cross sections for reactions of supernova 
neutrinos on $^{16}$O and $^{208}$Pb target nuclei  
as functions of the temperature and chemical potential.

The results for the test cases are in good agreement with the few available 
data, and with the cross sections calculated in the shell model for reactions 
on light nuclei. The advantage of the RHB plus PN-RQRPA model over the 
shell model approach is, of course, the possibility of performing calculations 
for reactions on heavier nuclei and in regions of nuclei far from stability. 
The differences between the results of various RPA-based calculations, 
especially in heavier nuclei and at low neutrino energies, can most probably 
be attributed to the different effective interactions used in modeling the structure 
of target nuclei, but they also indicate that more detailed experimental and 
theoretical studies of the transitions that contribute to the neutrino reaction
rates must be performed.

After the present test study, future applications of the RHB plus PN-RQRPA model 
will include a more extensive investigation of charged-current neutrino-nucleus 
reaction rates, as well as processes induced by neutral-current neutrino reactions, 
and neutrino nucleosynthesis.

%=========================================================================

\bigskip \bigskip
%%%%%%%%%%%%%%%%%%%%%%%%%%%%%%%%%%%%%%%%%%%%%%%%%%%%%%%%%%%%%%%%%%%%%

%==================================================================
\leftline{\bf ACKNOWLEDGMENTS}
\noindent
This work has been supported in part by MZOS - 
project 1191005-1010, and by the DFG Cluster of Excellence
   \textquotedblleft Origin and Structure of the Universe\textquotedblright .
%==================================================================
%\newpage

\newpage

%-----------------------------------------------------------------------
%-----------------------------------------------------------------------

%-----------------------------------------------------------------------
%-----------------------------------------------------------------------
%-----------------------------------------------------------------------
%-----------------------------------------------------------------------

%-----------------------------------------------------------------------
%---------------------------------------------------------
%---------------------------------------------------------
%-----------------------------------------------------------------------

%-----------------------------------------------------------------------
%------------------------------------------------------------------------
%------------------------------------------------------------------------
%------------------------------------------------------------------------

%-----------------------------------------------------------------------------------------------------------------------
\begin{table}
\caption{Flux-averaged neutrino-nucleus cross sections for
the $^{12}$C target nucleus. The results of several shell-model and 
QRPA based calculations, including the present 
PN-RQRPA(DD-ME2), are compared with available data.}
\begin{center}
%\begin{indented}
%\lineup
\item[]\begin{tabular}{@{}*{3}{l}}
\hline                        
$ $&$(\nu_{\mu},\mu^-)$&$ (\nu_{e},e^-)$\cr
$ $&$\langle \sigma \rangle (10^{-40}cm^2)$&$\langle\sigma\rangle (10^{-42} cm^2)$\cr 
\hline
RPA(SIII)~\cite{Vol.00}&19.23&55.10\cr
QRPA(SIII)~\cite{Vol.00}&20.29&52.0\cr 
CRPA(WS+LM)~\cite{Kol.99} &18.18&19.28\cr
PQRPA(PIII)~\cite{Krm.05}& &17.54 \cr
RPA~\cite{AAS.06}& & 13.60\cr
RPA(SGII)~\cite{Aue.97}&13.5&12.9\cr
RPA(SIII)~\cite{Aue.97}&14.5&16.5\cr
SM(WS(0+1+2+3)$\hbar\omega$)~\cite{Hay.00}&13.2&12.3\cr
SM(0$\hbar\omega~\times$0.64)~\cite{Aue.02}&19.2 &15.1\cr
SM(HF(0+1+2+3)$\hbar\omega$)~\cite{Vol.00}&15.18&16.42\cr
PN-RQRPA(DD-ME2) &19.59&12.14\cr
\hline
Exp.(KARMEN)~\cite{Bod.94}& &14.0$\pm$1.2\cr
Exp.(LSND)~\cite{LSND.98,LSND.01}&12.4$\pm$0.3$\pm$1.8 & 13.2$\pm$0.4$\pm$0.6\cr
Exp.(LAMPF)~\cite{Kra.92}& &14.1$\pm$2.3\cr 
\hline
\end{tabular}
\end{center}
\label{table1}
\end{table}
%------------------------------------------------------------------------------------------------------------------------

%------------------------------------------------------------------------------------------------------------------------
\begin{table}
\caption{Flux-averaged cross sections for the $\nu_e$ reaction on
$^{16}$O, $^{56}$Fe, and $^{208}$Pb target nuclei.}
\begin{center}
%\begin{indented}
%\lineup
\item[]\begin{tabular}{@{}*{4}{l}}
\hline
%\br                              
 &$^{16}$O($\nu_{e},e^-$)$^{16}$F&$^{56}$Fe($\nu_{e},e^-$)$^{56}$Co&$^{208}$Pb($\nu_{e},e^-$)$^{208}$Bi\cr
$ $&$\langle\sigma\rangle (10^{-42} cm^2)$&$\langle\sigma\rangle (10^{-42} cm^2)$&$\langle\sigma\rangle (10^{-42} cm^2)$ \cr 
%\mr
\hline
SM(0$\hbar\omega~\times$0.64)~\cite{Aue.02} &10.8 & &\cr
TM~\cite{Min.02} &  &214&\cr
TDA(SKIII)~\cite{Suz.03}& & &2954,3204\cr
RPA~\cite{AAS.06}&14.55&277&2643\cr
CRPA(WS+LM)~\cite{Kol.01}&&240&3620\cr
(Q)RPA(SIII,SGII)&16.90,17.20~\cite{Aue.97}&352~\cite{Laz.07}&4439~\cite{Vol.02}\cr
PN-RQRPA(DD-ME2)&13.18&140&2812\cr
\hline
Exp.(KARMEN)~\cite{Mas.98,Kol.01}& &256$\pm$108$\pm$43\cr
\hline
%Mean Theory&15.45&244.6&3396\cr
%\br
\end{tabular}
%\end{indented}
\end{center}
\label{table2}
\end{table}
%----------------------------------------------------------------------------------------------------------------------

\end{document}